\documentclass[namedreferences]{SolarPhysics}
\usepackage{solaheader}    
\usepackage[optionalrh]{spr-sola-addons} 

\usepackage{graphicx}        
\usepackage{url}             
\usepackage{marvosym}
\usepackage{times}

\newcommand{\etal}{{\it et al.}}

\newcommand{\aap}{    {\it Astron. Astrophys.}}

\newcommand{\apj}{    {\it Astrophys. J.}}

\newcommand{\solphys}{{\it Solar Phys.}}

\begin{document}

\begin{article}

\begin{opening}

\title{Spectral Inversion of Multi-Line Full-Disk Observations of Quiet
    Sun Magnetic Fields}
\author{H.~\surname{Balthasar}\sep M.L.~\surname{Demidov}}
\runningauthor{H.\ Balthasar, M.L.\ Demidov}
\runningtitle{Inversion of Multi-Line Observations of the Quiet Sun
    Magnetic Fields}
\institute{
    H.\ Balthasar ({\large \Letter})\\
        \mbox{Leibniz-Institut f\"ur Astrophysik Potsdam (AIP),}
        \mbox{An der Sternwarte 16, 14482 Potsdam, Germany}\\
        email: \url{hbalthasar@aip.de}\\\smallskip
    M.L.\ Demidov\\
        \mbox{Institute of Solar-Terrestrial Physics, Siberian Branch,}\\
        \mbox{Russian Academy of Sciences, 664033 Irkutsk, P.O.\ Box 291,
        Russia}\\
        email: \url{demid@iszf.irk.ru}}

\begin{abstract}
Spectral inversion codes are powerful tools to analyze spectropolarimetric
observations, and they provide important diagnostics of solar magnetic fields.
Inversion codes differ by numerical procedures, approximations of the
atmospheric model, and description of radiative transfer. \textit{Stokes
Inversion based on Response functions} (SIR) is an implementation widely used by
the solar physics community. It allows to work with different atmospheric
components, where  gradients of different physical parameters are possible,
\textit{e.g.}, magnetic field strength and velocities. The spectropolarimetric
full-disk observations were carried out with the Stokesmeter of the
\textit{Solar Telescope for Operative Predictions} (STOP) 
at the Sayan Observatory on 3~February 2009,
when neither an active region nor any other extended flux concentration was
present on the Sun. In this study of quiet Sun magnetic fields, we apply the SIR
code simultaneously to 15 spectral lines. 
A tendency is found that weaker magnetic field strengths occur closer to the 
limb. We explain this finding by the fact that close to the limb, we are more 
sensitive to higher altitudes in an expanding flux tube, where the field strength 
should be smaller since the magnetic flux is conserved with height.
Typically, the inversions deliver two populations 
of magnetic elements: (1) high magnetic field strengths (1500--2000~G) and
high temperatures (5500--6500~K) and (2) weak magnetic fields (50--150~G) and
low temperatures (5000--5300~K).
\end{abstract}
\keywords{Magnetic fields -- Photosphere\authorsep Spectral Line -- Intensity
    and Diagnostics\authorsep Polarization -- Optical\authorsep Center-Limb
    Observations}
\end{opening}

\section{Introduction\label{S-Introduction}}

The small-scale organization of solar magnetic fields still holds many
mysteries. Even with observations from space, where the Japanese \textit{Hinode}
mission (\opencite{Suematsu-etal08}) now routinely provides spectropolarimetric
data in the two Fe\,\textsc{i} $\lambda 630.15$~nm and $\lambda 630.25$~nm lines
with high angular resolution and high polarimetric precision, the fundamental
spatial scales remain just beyond our grasp. Various investigations based on
\textit{Hinode} data, however, demonstrate the rapid advances in understanding
solar small-scale magnetic fields (\opencite{Ishikawa}; \opencite{Kontogiannis};
\opencite{Manzo Sainz}; \opencite{ThorntonParnell}). The next generation of
ground-based telescopes with apertures larger than 1-meter diameter and future
space missions such as the \textit{Solar Orbiter} will certainly provide
magnetic field data with even higher angular resolution. But there are some
indications that the smallest building blocks of solar magnetic fields might be
as small as 10~km (\opencite{Stenflo2010}, \citeyear{Stenflo2011}). This would
be a challenge even for the largest telescopes, which are currently on the
drawing boards.

The \textit{Helioseismic and Magnetic Imager} (HMI) on board of the
\textit{Solar Dynamics Observatory} (SDO, \opencite{Norton-etal06};
\opencite{Borrero-etal10}) opens other avenues to understanding solar magnetism.
Full-Stokes observations are carried out with moderate spectral resolution in
the Fe\,\textsc{i} $\lambda 617.33$~nm line with a high cadence of 45~s covering
the entire solar disk with a spatial resolution of about one second of arc.
Besides exploiting these data in the context of helioseismology, they allow us
to study the dynamics as well as the long-term evolution of solar magnetic
fields. Finally, 3-D magneto-hydrodynamic (MHD) simulations (\textit{e.g.},
\opencite{Danilovic-etal10}) provide another perspective elucidating the nature
of solar small-scale magnetic fields.

In this study, we use a complementary approach to extract the inner (hidden)
properties of magnetic fields contained inside the resolution element of the
Stokesmeter of the \textit{Solar Telescope for Operative Predictions} (STOP)
at the \textit{Sayan Solar Observatory} near Irkutsk, Russia.
In a two-components approach, where the magnetic and
non-magnetic components have different filling factors within the resolution
elements, as many spectral lines as possible are necessary for reliable magnetic
field diagnostics (\opencite{DemBalt11}). We studied the center-to-limb variation (CLV) 
of quiet Sun magnetic fields, \textit{i.e.}, the dependence of physical parameters
on the cosine of the heliocentric angle $\mu = \cos(\vartheta)$, 
using the code \textit{Stokes Inversion based on Response functions} (SIR) 
(\opencite{SIR-92}), where multi-line diagnostics and the
two-components approach can be implemented in a straightforward manner. The
present investigation combines three important aspects: (1) we explore the
Stokesmeter observations (Stokes $I$ and $V/I_{c}$ profiles) in 15 spectral
lines simultaneously, (2) we analyze observations covering practically the whole
range of heliocentric angles $\vartheta$, and (3) we analyze signals of
extremely small degrees of polarization. Our investigation complements
high-resolution observations and potentially links them to spatially unresolved
stellar observations.

\section{Observations\label{S-Observations}}

This study is an extension of our previous investigations (\opencite{DemBalt09},
\citeyear{DemBalt11}), where we analyzed Stokesmeter data of 15 spectral lines,
which were obtained with the STOP telescope (\opencite{Demidov-etal02}) in February 2009. A
detailed description of the instrument, the observing procedure, and examples of
Stokes profiles, as well as the line parameters can be found in these
publications. Nominally, spectra are obtained for areas covering
$100^{\prime\prime}$ on a raster with an equidistant grid spacing of
$91^{\prime\prime}$ covering the entire solar disk. In this study, this spacing
was kept but the areas were reduced to $10^{\prime\prime} \times 10^{\prime\prime}$. 
Spectra for 325 of
such equidistantly spaced pixels were recorded with an integration time of 8~s.
Thus, a raster scan of the entire solar disk takes about 70~min including the time 
to position the telescope and the polarimeter.
The noise level in the continuum is used to estimate the polarimetric accuracy.
We find an \textit{rms}-value of { } $0.5 \times 10^{-4}$.

\begin{figure}[t]
\centerline{\hspace*{0.015\textwidth}
\includegraphics[width=0.48\textwidth,clip=]{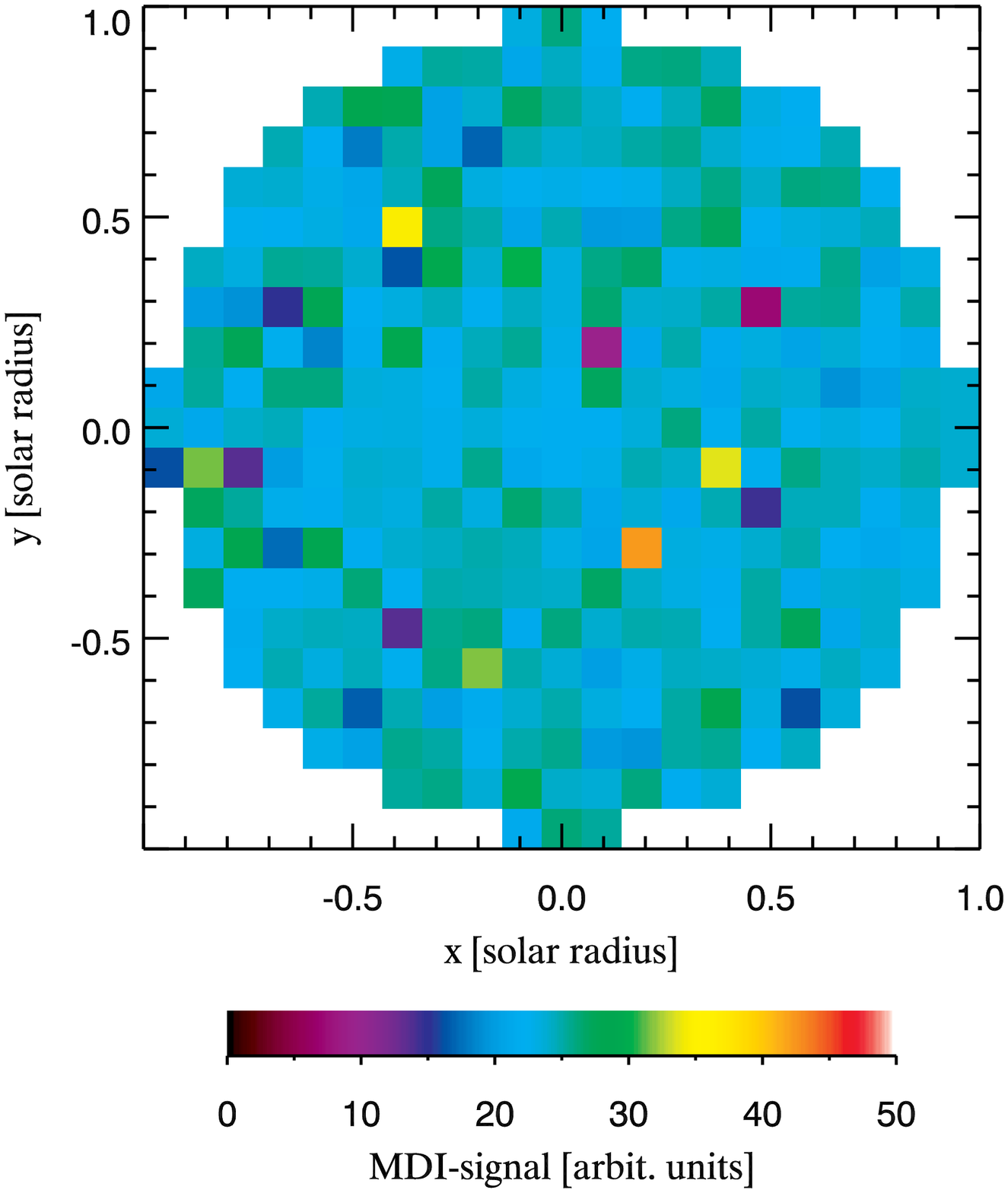}
\hspace*{0.02\textwidth}
\includegraphics[width=0.48\textwidth,clip=]{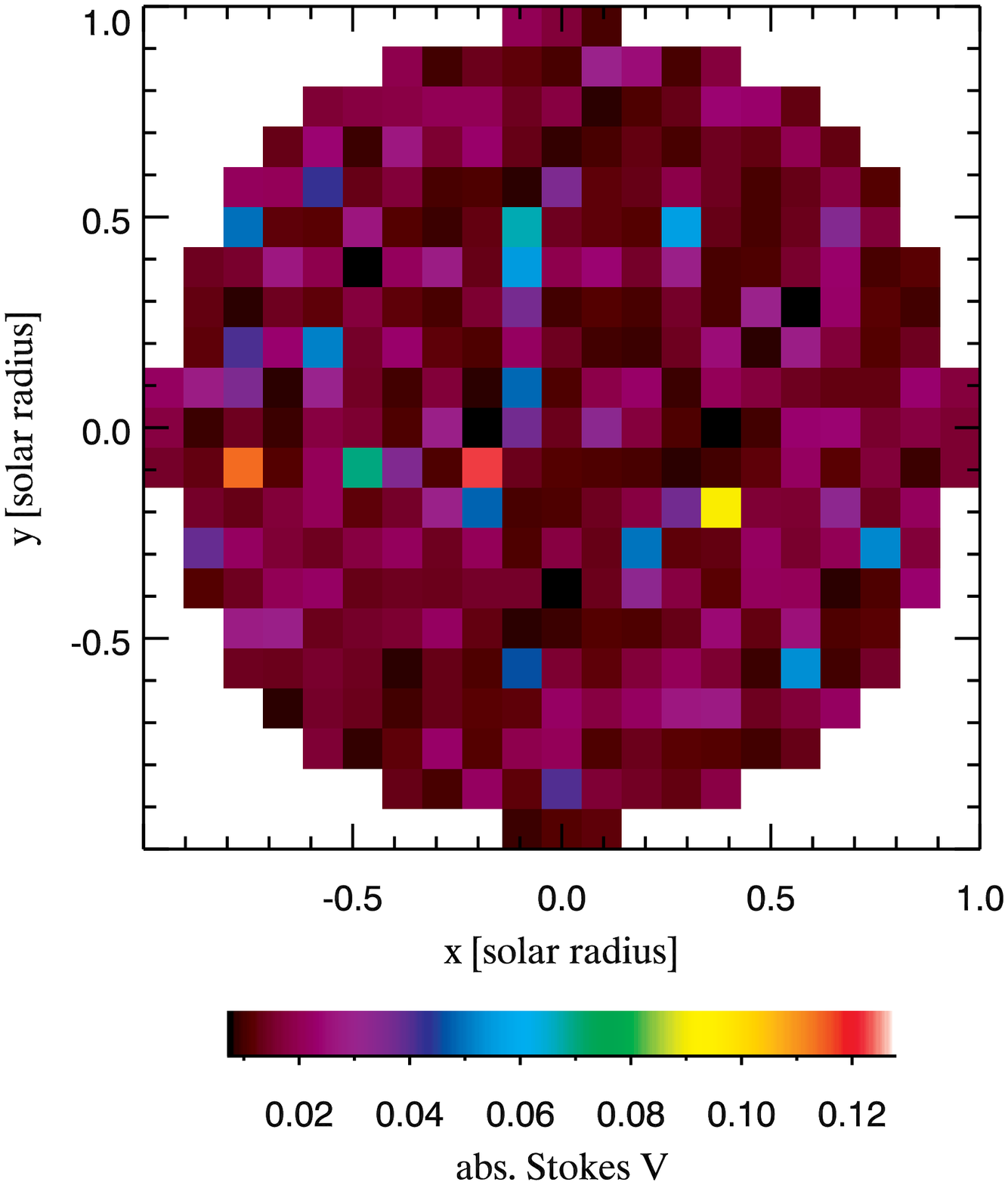}
}
\centerline{\hspace*{0.015\textwidth}
\includegraphics[width=0.48\textwidth,clip=]{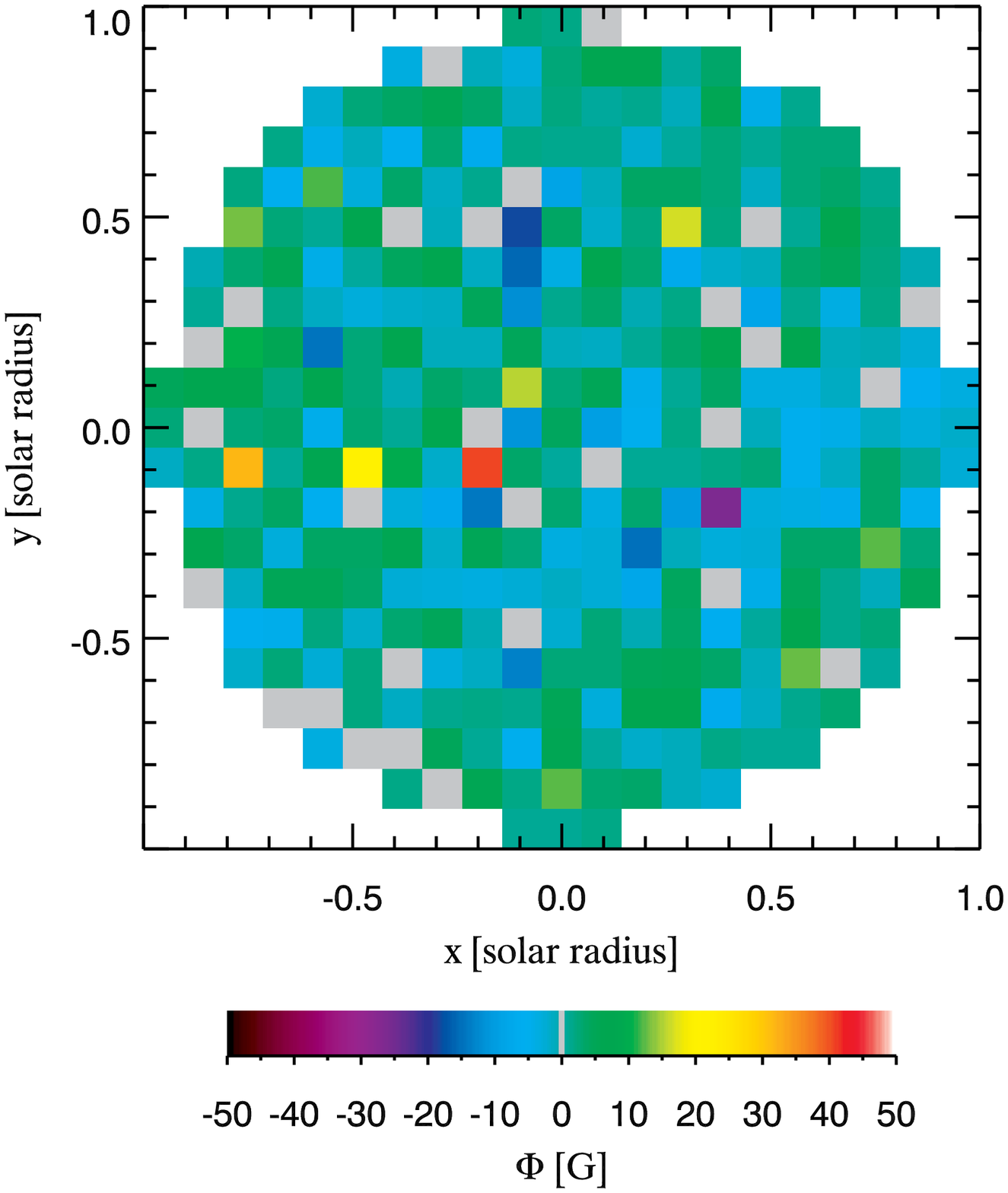}
\hspace*{0.02\textwidth}
\includegraphics[width=0.48\textwidth,clip=]{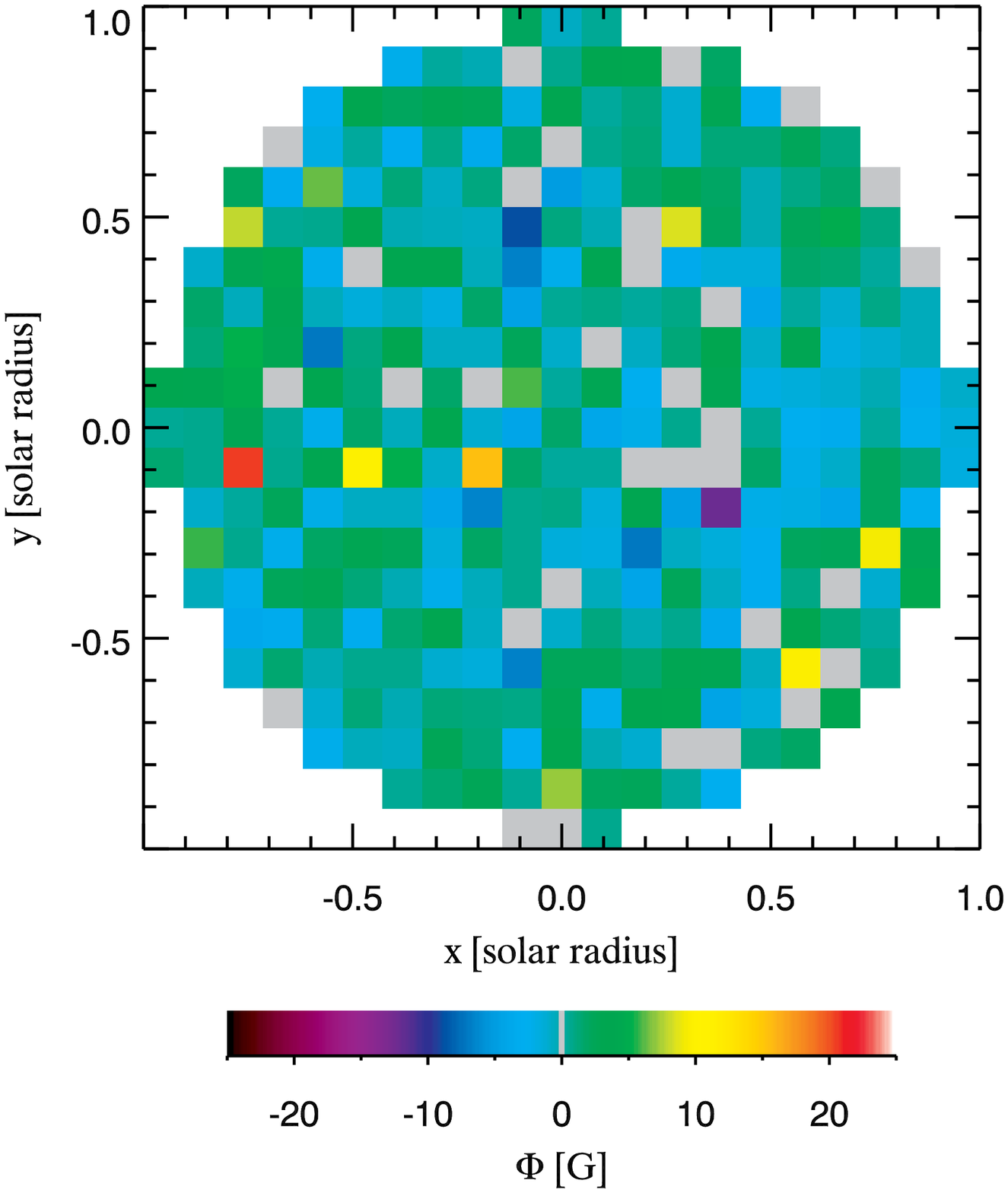}
}
\caption{Averaged MDI-magnetogram (upper left) and
integrated absolute values of the Stokes $V/I_{c}$ parameter $A$ (upper right).
The bottom row shows full-disk magnetograms observed simultaneously in the 
Fe\,\textsc{i}
    $\lambda 523.3$~nm (left) and $\lambda 525.0$~nm (right) lines on 3~February
    2009. Each pixel corresponds to $10^{\prime\prime} \times
    10^{\prime\prime}$. Note that the size of the pixels is not to scale.}
\label{F-2mag}
\end{figure}

Solar activity was extremely low during the observations in February 2009. No
sunspot was present during the first half of the month and only a small emerging
flux region received an NOAA number on 11~February 2009. The appearance of the
quiet Sun magnetic fields resembles that of a salt-and-pepper pattern, which
covered the entire solar surface. We selected 3~February 2009 for this study
because on this day, the seeing conditions were good and a large number of
pixels exceeded the noise threshold of the Stokes $V / I_c$ signal. To validate
our observations, we averaged a magnetogram of the \textit{Michelson
Doppler Imager} (MDI) on board the \textit{SOlar and Heliospheric Observatory
(SOHO)} using the same grid  and size of pixels as in the STOP Stokesmeter
observations. The used magnetogram was obtained on the same day, but some hours 
later. The visual appearance of the two magnetic field rasters is
surprisingly similar.

In this study, we distinguish strictly between the magnetic flux density as we
derive it from our classical method (\opencite{Demidov-etal08}) and
the magnetic field strength $B$ we obtain for the magnetic component from
the SIR-inversions. In contrast to our previous studies, we use now the 
symbol $\Phi$ for the magnetic flux density. 
Magnetic features in the quiet Sun are small and do not fill our resolution
element, thus we deal with flux densities. From the inversion, we derive also
a filling factor and therefore we consider the inversion result as magnetic field 
strength.
The magnetic flux densities $\Phi_{523.3}$ and  $\Phi_{525.0}$ are shown in
Figure \ref{F-2mag} for the pair of Fe\,\textsc{i} $\lambda 523.3$~nm and
$\lambda 525.0$~nm lines. There are two pixels on both maps with remarkably high
flux densities -- one near the eastern limb at $\mu = 0.65$,
and another close to disk center at $\mu = 0.98$. Just
comparing the flux densities of these central and limb pixels in the two
Fe\,\textsc{i} lines provides a first hint for a strong CLV of the magnetic
flux density ratio $R$. The flux densities are $\Phi_{523.3} = 41.0$~G and
$\Phi_{525.0} = 15.5$~G for the central and $\Phi_{523.3} = 31.8$~G and
$\Phi_{525.0} = 20.6$~G for the limb pixels, respectively. This is in good
agreement with the strong CLV of the magnetic flux density ratio $R(\mu) =
\Phi_{523.3}(\mu) / \Phi_{525.0}(\mu) = 1.74 - 2.43\mu + 3.43\mu^2$ with a
disk-averaged ratio of $\bar{R}=1.97 \pm 0.02$, which was measured by
\inlinecite{DemBalt09}.

We introduce now the parameter $A$ which is the integrated absolute value of 
Stokes $V/I_{c}$ divided by 15 (the number of lines), shown in Figure~\ref{F-2mag}.
This parameter was used as a measure of the polarization signal because the 
measured flux densities in different lines vary from pixel to
pixel. The majority of pixels has polarization signals, which are too weak to
yield reliable inversion results. Pixels with $A$ below
0.025 were removed from further analysis. 
According to visual inspection, polarization signals are above the noise level
if $A$ exceeds this threshold.
This threshold corresponds to a magnetic flux density of roughly 3.5~G from the 
Fe\,\textsc{i}$\lambda 525.0$~nm line.
In addition, we did not consider pixels closer to the limb than $\mu = 0.6$.
A total of 37 pixels surpassed these criteria
and form the basis for subsequent analysis.

In a first step, SIR was applied to these pixels. The starting model was
basically the same as in \inlinecite{DemBalt11}. A pixel was considered to be
composed of non-magnetic (\textit{Harvard Smithsonian Reference Atmosphere}, HSRA, 
\opencite{HSRA-71}) and magnetic components
with a height-independent filling factor $f$ as a free parameter. The magnetic
component has the temperature stratification $T(\tau)$ of the model of
\inlinecite{solanpla}, but a different stratification for the magnetic field.
From this initial guess, temperatures were calculated with five nodes 
in optical depth $\tau$. In
contrast to \inlinecite{DemBalt11}, the magnetic field strength $B$ was
considered to be constant with depth and oriented vertically with respect to
the solar surface. The angle $\gamma$ between the line-of-sight (LOS) and the
direction of the field lines coincides either with $\vartheta$  or with
180$^\circ - \vartheta$ depending on the sign of Stokes $V/I_{c}$.
Intensity spectra are normalized for the inversions in a way that they 
follow the CLV of Equation~10 in the article of \inlinecite{Pierce77},
since we do not know the exact variation of the atmospheric transparency 
during our observations.

\section{Results\label{Results}}
\begin{figure}[t]    
    \centerline{
     \includegraphics[width=115mm,height=140mm,clip=]{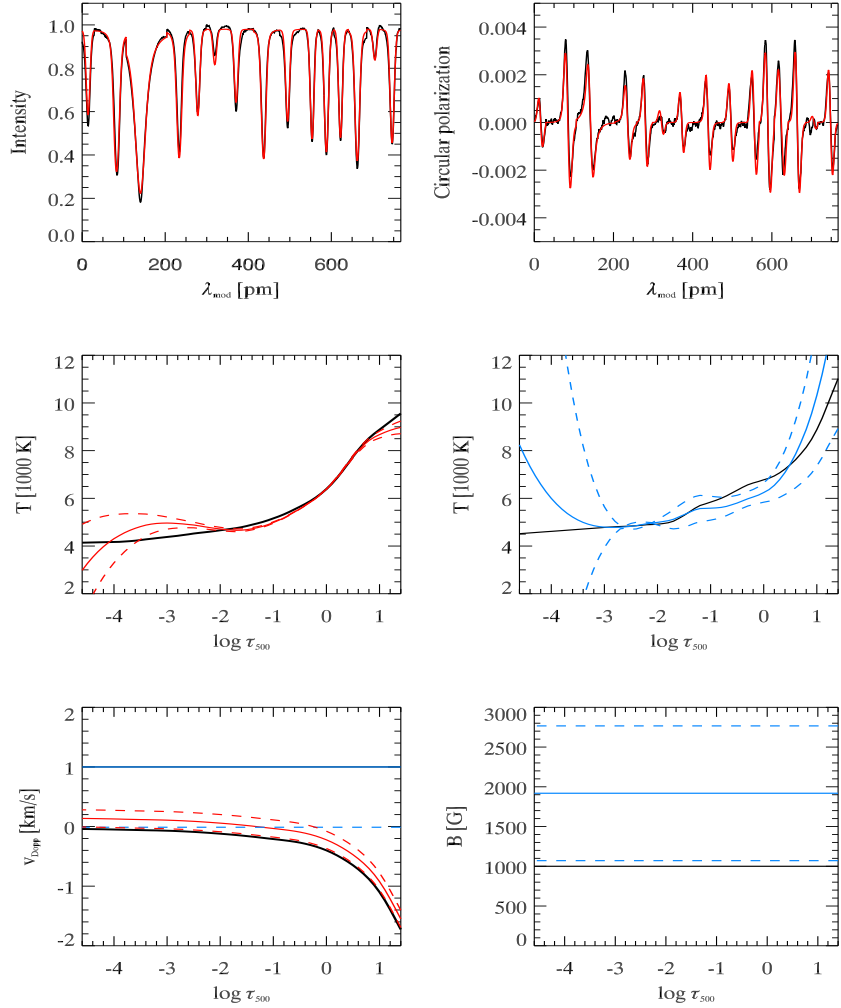}
  }
\caption{Comparison of observed (black) and inverted (red) Stokes $I$ 
    (upper-left) and Stokes
    $V/I_{c}$ (upper-right) profiles. The inversion results are shown for one
    pixel close to disk center : temperatures $T$ for the magnetic and
    non-magnetic (middle-left) and magnetic (middle-right) components, 
    Doppler velocities $v_\mathrm{Dopp}$ (bottom-left), and magnetic field $B$ 
    (lower-right), Blue and red lines indicate the
    magnetic and non-magnetic components, respectively.
    The starting model is represented by solid black
    lines. Error limits of the inversion are
    given by the dashed lines. The wavelength scale is not continuous,
    \textit{i.e.}, the spectral regions of interest are first extracted and then
    collated as in Demidov and Balthasar (2012).}
\label{F-Horst ex}
\end{figure}

As an example, a comparison of observed and inverted spectral line profiles is
shown in Figure~\ref{F-Horst ex}, which also depicts the parameters of the
starting model and the inversion results. The magnetic filling factor in this 
special case is $0.016 \pm 0.006$. The inverted profiles for the magnetic
and non-magnetic components agree very well with the observations. In quiet Sun
regions, we assumed that hot rising granules contribute more in deep atmospheric
layers. Therefore, only a shift of the entire Doppler velocity curve was allowed
in the inversions. The displayed error ranges are delivered by the SIR code.
Temperatures become rather unreliable in very high layers,
and the steep temperature increase of the magnetic component above
$\log\tau_{500} = -3.5$ is most likely not realistic.

\begin{figure}[t]    
    \centerline{\hspace*{0.015\textwidth}
                \includegraphics[width=0.48\textwidth,clip=]{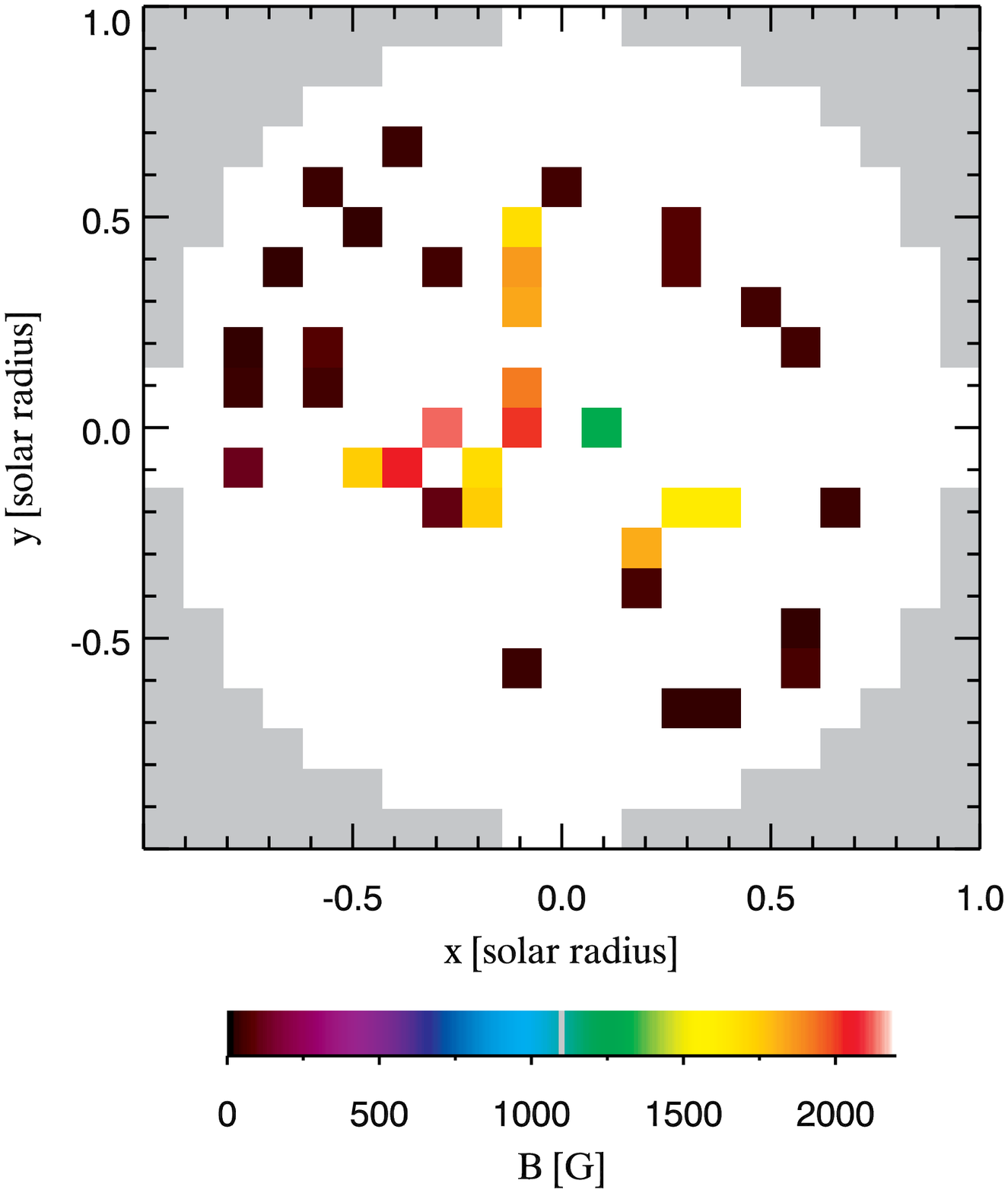}
               \hspace*{0.02\textwidth}
                \includegraphics[width=0.48\textwidth,clip=]{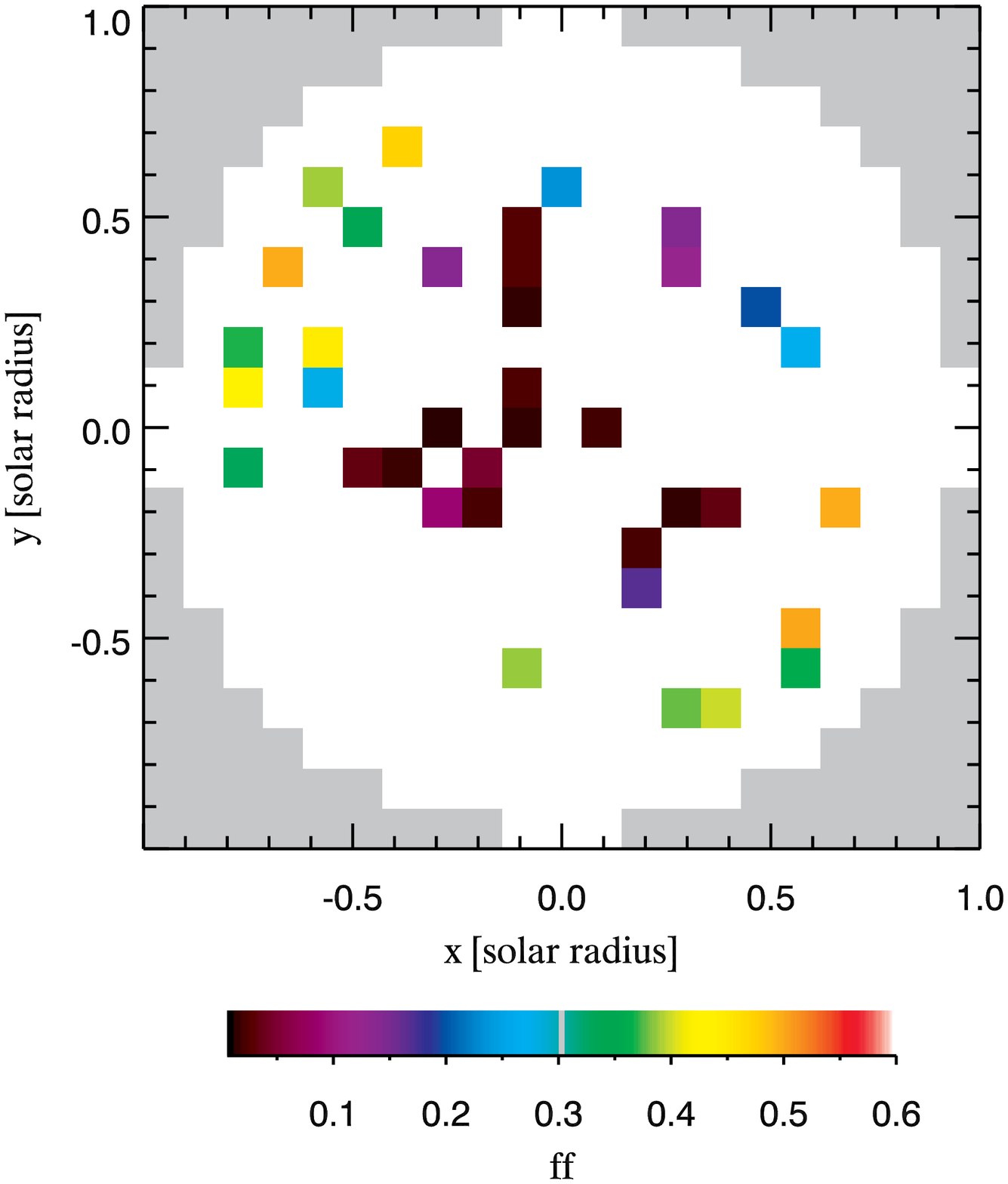}
              }
    \centerline{\hspace*{0.015\textwidth}
               \includegraphics[width=0.48\textwidth,clip=]{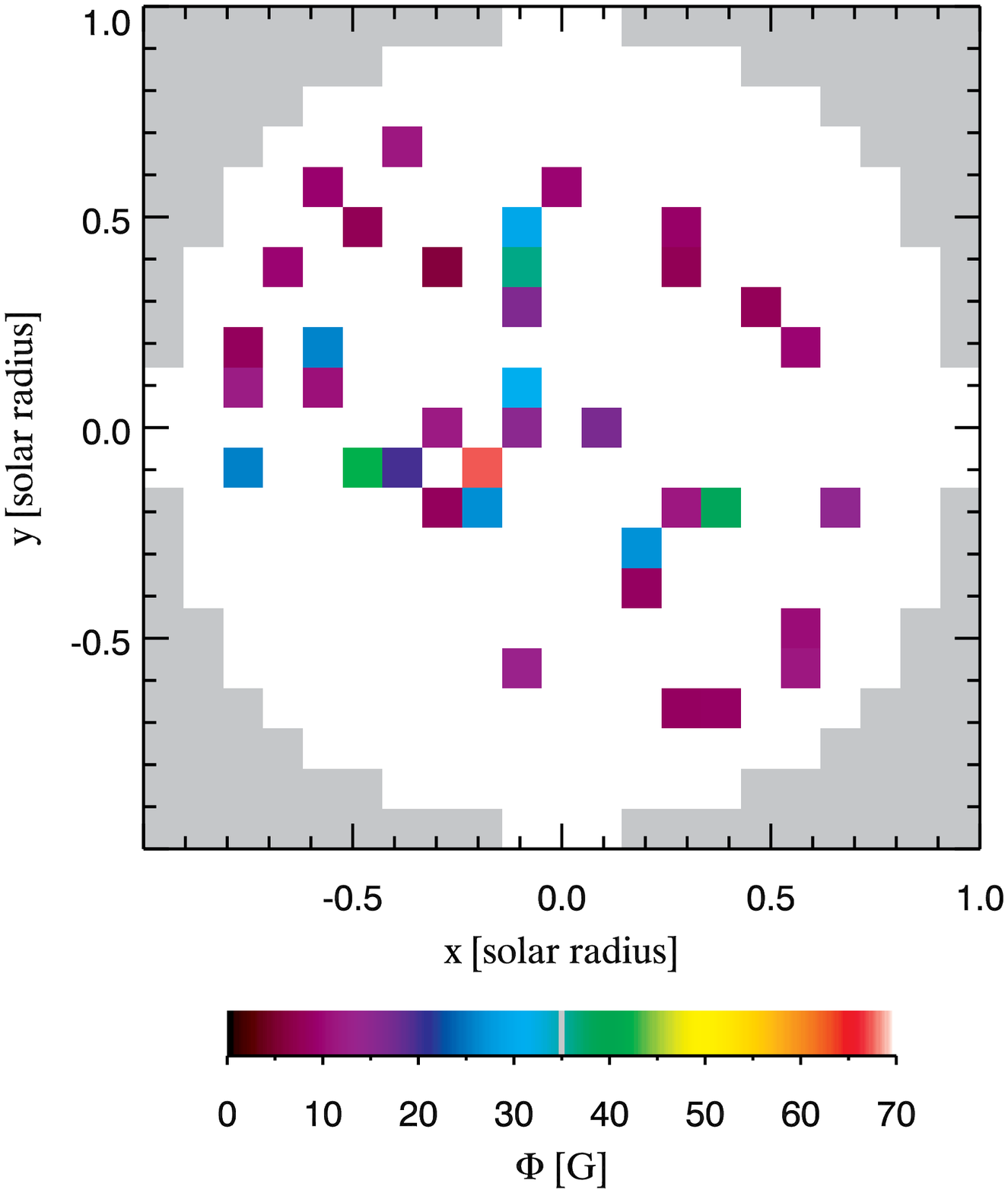}
               \hspace*{0.02\textwidth}
                \includegraphics[width=0.48\textwidth,clip=]{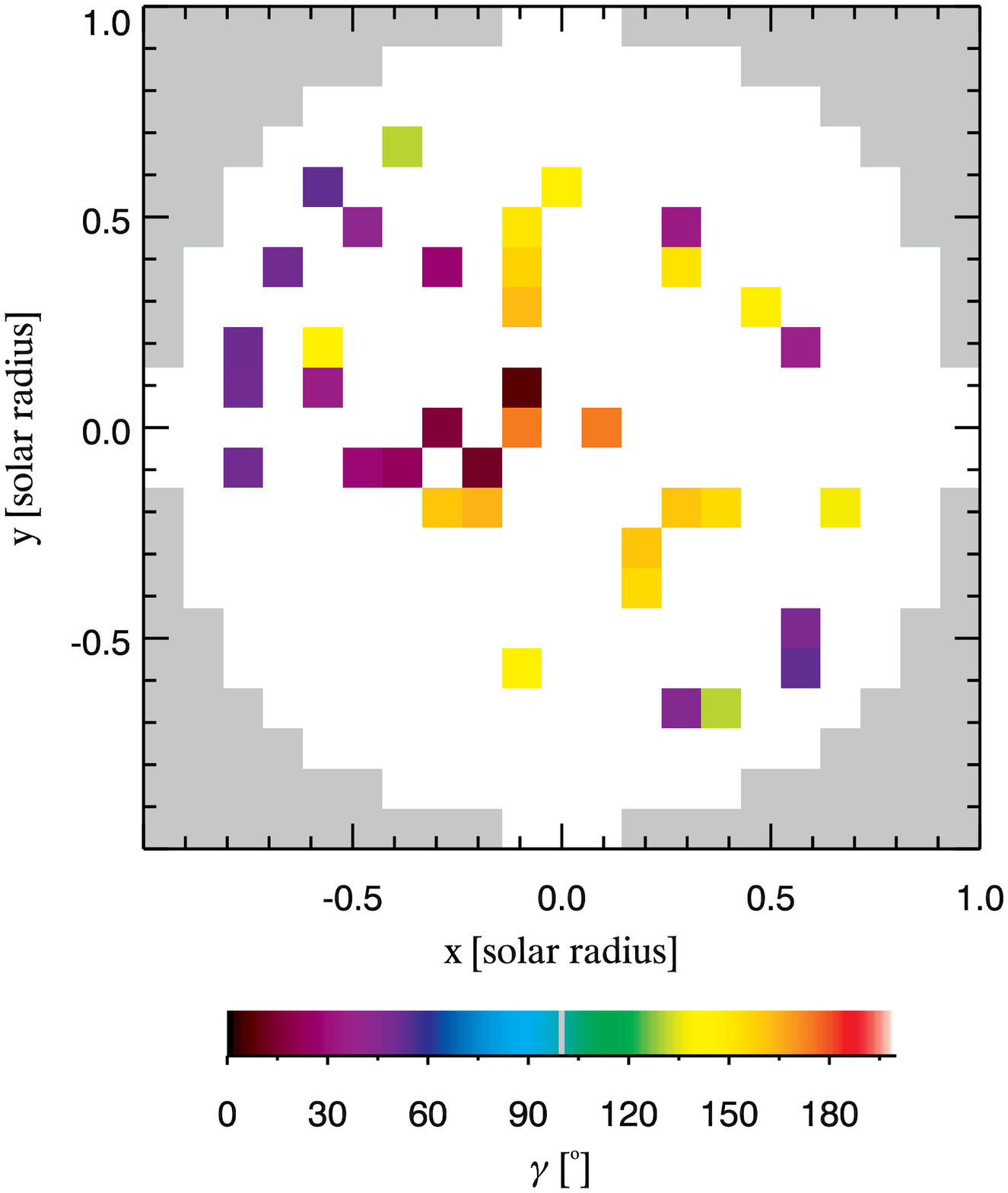}
               }
\caption{Inversion results for the magnetic field strength $B$ (upper left), 
    filling factor $f$ (upper right), and the LOS magnetic flux density 
   $\Phi$ (lower left). The angle $\gamma$ in the selected pixels is displayed
   in the lower right panel.
     }
\label{F-Binv and FF}
\end{figure}

The SIR code was applied to all pixels with Stokes $V/I_c$ signals above the
noise threshold defined in Section~\ref{S-Observations}. The inverted magnetic
field strengths $B$ and filling factors $f$ for these pixels are presented in
the upper left and right panels of Figure~\ref{F-Binv and FF}, respectively. 
The inversions yield two populations with a distinct dependence on the heliographic
angle $\vartheta$: (1) pixels with kilo-Gauss strengths (larger than 1200\,G) and very
small filling factors (0.005 -- 0.041) in the central part of the solar disk and
(2) pixels with rather weak field strengths 
(less than 200\,G) and large filling factors (up to 0.5) closer to the
solar limb. All strong field cases occur at $\mu > 0.85$ and all weak field 
cases at $\mu < 0.95$.

Knowing the inverted magnetic field strengths $B$ and filling factors $f$ of the
magnetic elements, we calculated the LOS flux density $\Phi = B \times \cos
\gamma \times f$, where $\gamma$ denotes the angle between the LOS and the
magnetic field, which is assumed to be perpendicular to the solar surface
(lower right panel in Figure~\ref{F-Binv and FF}). The
distribution of the LOS flux density $\Phi$ is also shown in 
Figure~\ref{F-Binv and FF}.
Pixels with high LOS flux densities are not necessarily associated
with high observed flux densities and inverted magnetic field strengths, mainly because of
different filling factors.

\begin{figure}[t]   
    \centerline{\hspace*{0.015\textwidth}
                \includegraphics[width=0.48\textwidth,clip=]{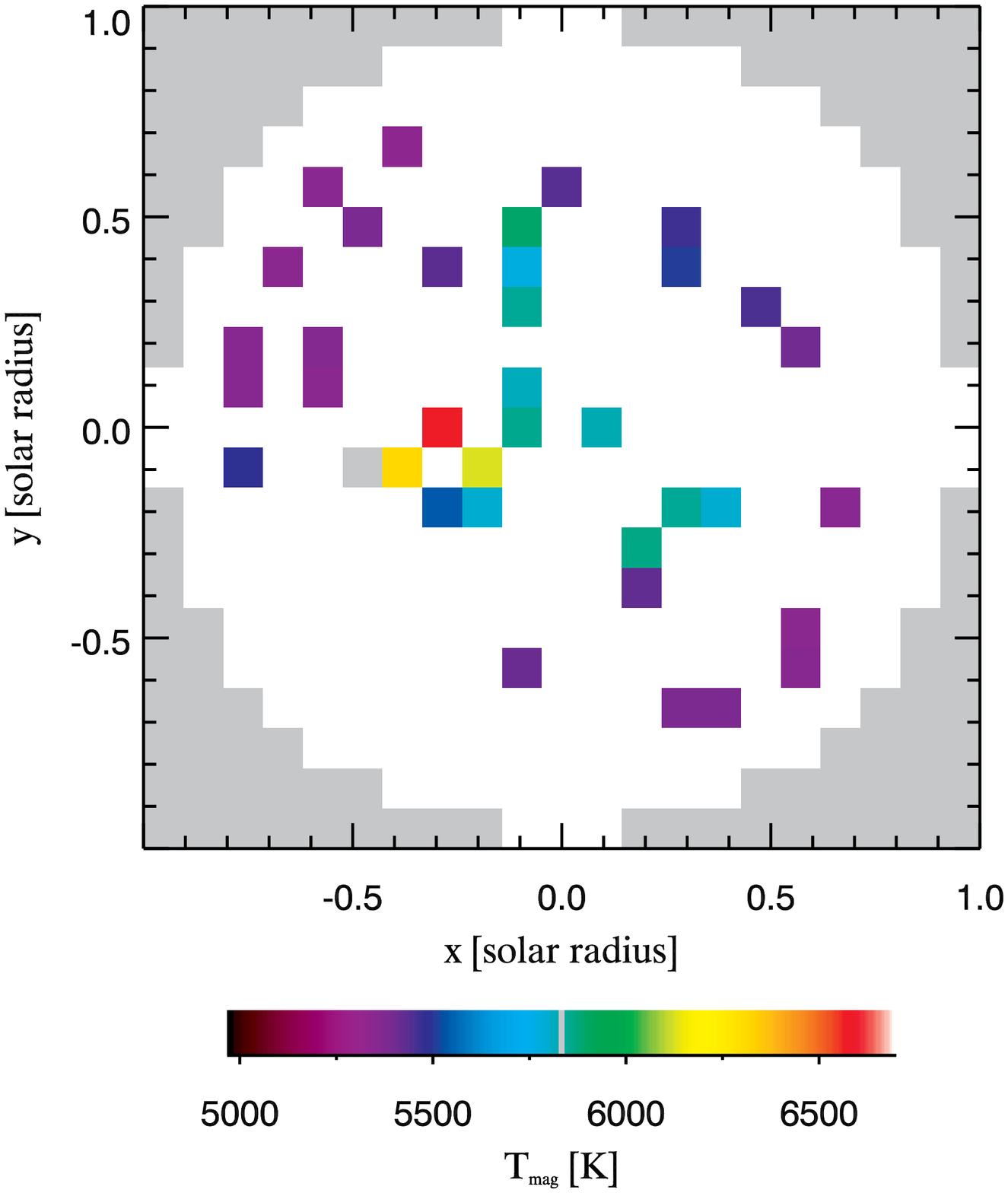}
               \hspace*{0.02\textwidth}
                \includegraphics[width=0.48\textwidth,clip=]{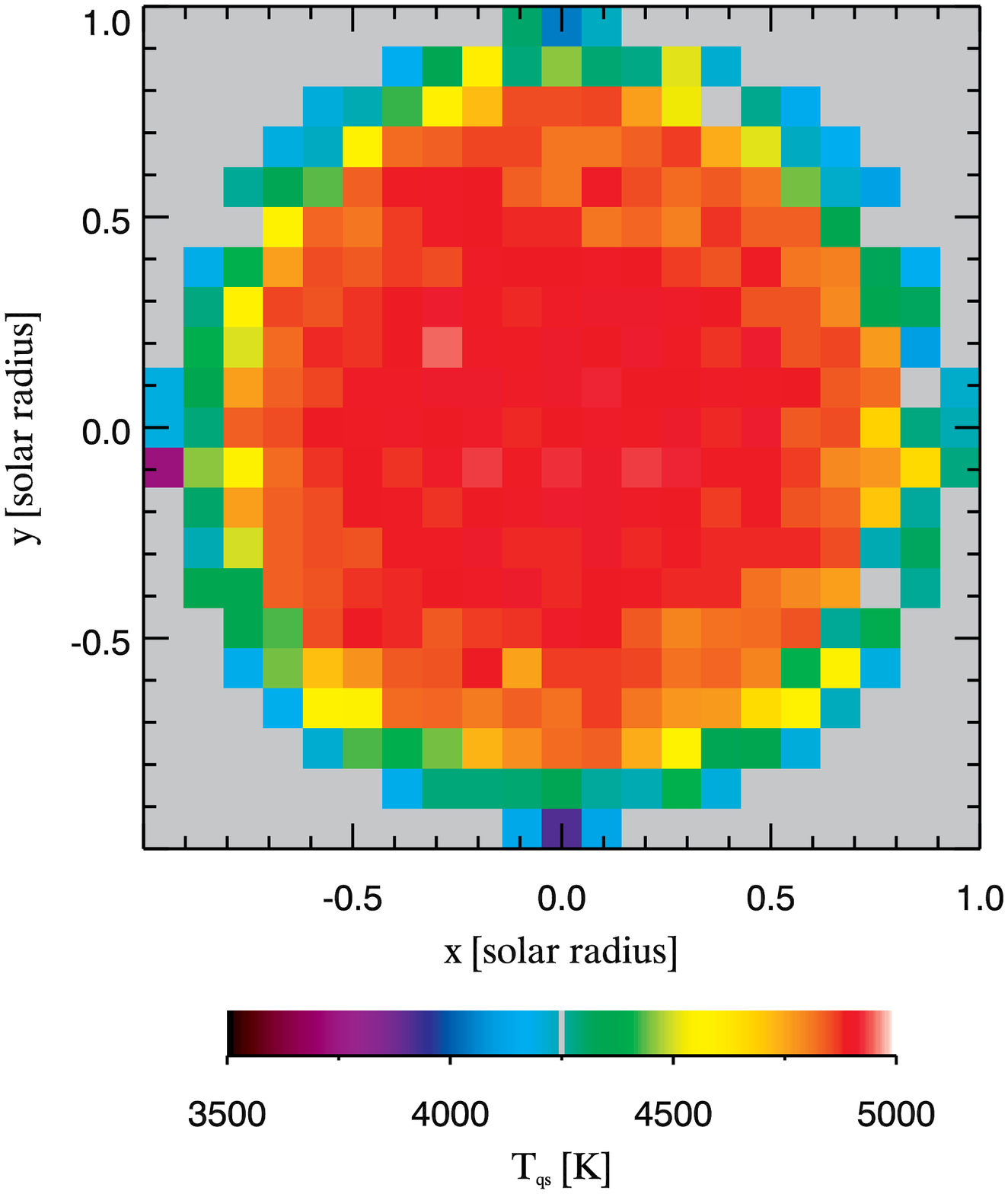}
               }
\caption{Inversion results for the temperature $T$ of the magnetic (left) and
    non-magnetic (right) components at an optical depth of $\log \tau_{500} =
    -1.5$.}
\label{F-Tme and Tqs}
\end{figure}

\begin{figure}[t]    
          \centerline{\hspace*{0.02\textwidth}
\includegraphics[width=0.80\textwidth,clip=]{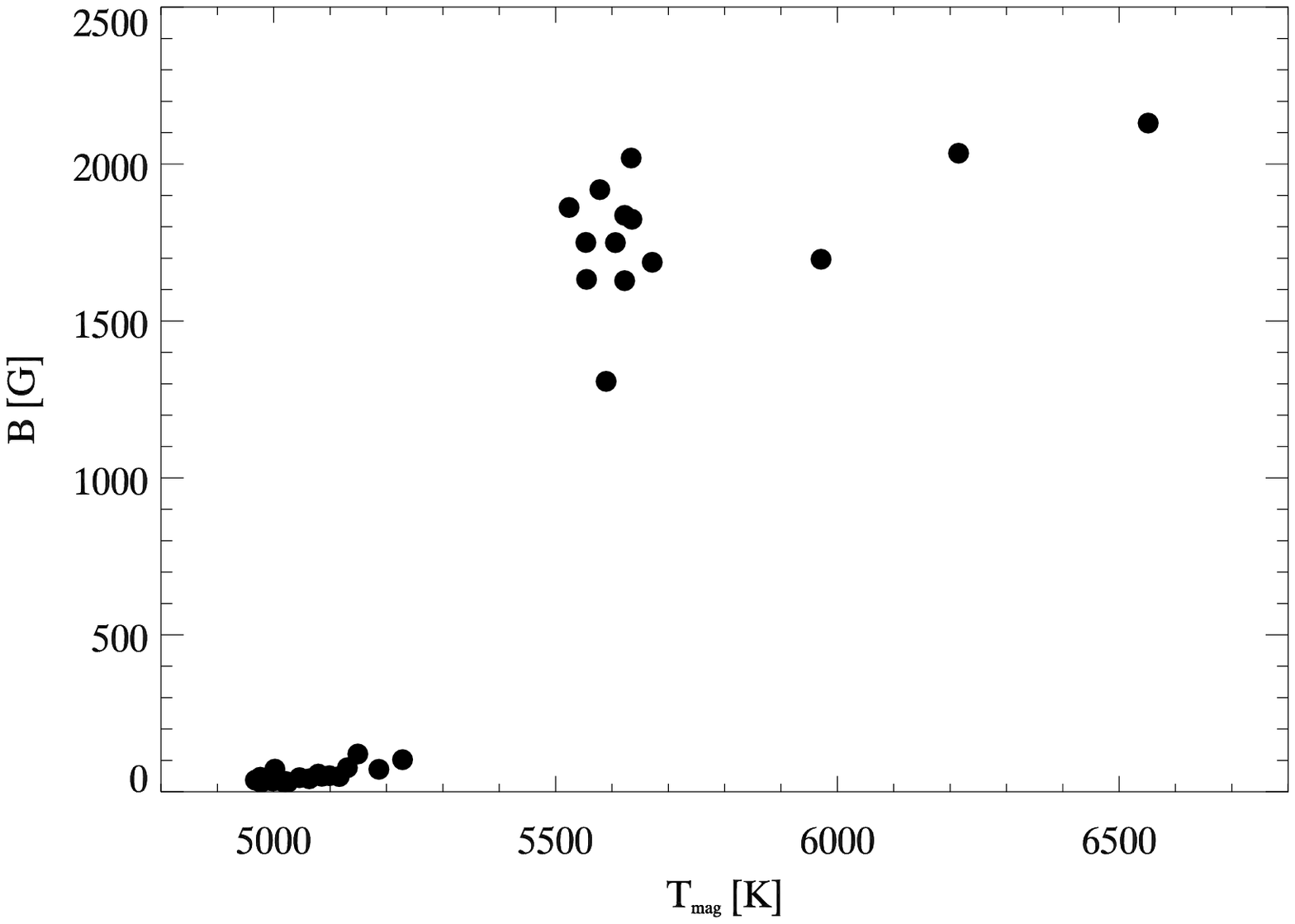}
                 }
\caption{Comparison of the inverted magnetic field strengths $B$ and temperature
    $T$, where $B$ is constant with depth and $T$ corresponds to an optical
    depth of $\log\tau_{500} = -1.5$.}
\label{F-B-T}
\end{figure}

The temperature can be derived for both the magnetic and non-magnetic components
(see Figure~ \ref{F-Tme and Tqs}). In both cases, the temperatures
$T_\mathrm{mag}$ and $T_\mathrm{qS}$ correspond to an optical depth of $\log
\tau_{500} = -1.5$. As long as the filling factor is not unity $f \not= 1$, the
quiet Sun temperature $T_\mathrm{qS}$ is known for all pixels on the solar disk.
The highest temperatures appear at disk center and the temperatures decrease
towards the limb as expected for a typical limb darkening profile. 
The dependence of $T_\mathrm{qS}$ on $\mu$ can be well described by this fit
$$ T_\mathrm{qS}(\mu) = 3030.03 + 3729.23\,\mu - 1835.61\,\mu^2. $$
The
temperature of the magnetic component $T_\mathrm{mag}$ adds another diagnostic
parameter for distinguishing between the two aforementioned populations. In the
first population (see Figure~\ref{F-Binv and FF}), strong magnetic fields
(1500--2000~G) are associated with high temperatures (5500--6500~K). In the
second population, weak fields (50--150~G) are accompanied by low temperatures
(5000--5300~K). The existence of two distinct populations also becomes clear,
when the magnetic field strengths $B$ are plotted against the temperatures $T$
(Figure~\ref{F-B-T}). There, the transition from the weak to the strong regime
does not occur smoothly but rather abruptly at a temperature of about 5300~K.

\begin{figure}[t]    
          \centerline{\hspace*{0.02\textwidth}
\includegraphics[width=0.80\textwidth,clip=]{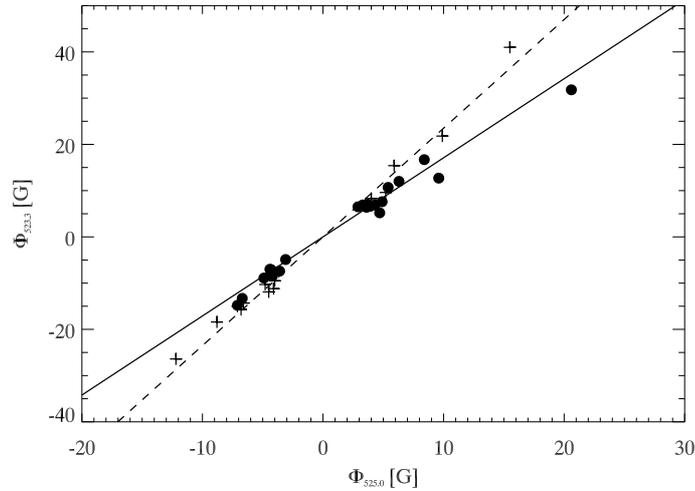}
                 }
\caption{Scatter plots for the magnetic flux densities $\phi$ of the two
    Fe\,\textsc{i} $\lambda 523.3$~nm and $\lambda 525.0~$nm lines. Linear
    regression models are shown for both the weak (bullets) and strong 
     (plusses) populations (see Figure~\ref{F-B-T}).}
\label{F-2regr}
\end{figure}

In previous investigations (\opencite{DemBalt09}, \citeyear{DemBalt11}), we
showed that for some combinations of spectral lines, the magnetic flux density
ratios $R$ have a rather strong CLV. The pair of Fe\,\textsc{i} $\lambda
523.3$~nm and $\lambda 525.0$~nm lines had the strongest CLV. Therefore, we
explored this combination of lines in more detail by computing the flux density
ratios separately for both populations (Figure~\ref{F-2regr}). Two
well-separated regression lines are clearly visible yielding flux density ratios of
$R_w = 1.71$ and $R_s = 2.35$ for the weak and strong populations, respectively.
Even a few abnormal pixels near disk center, which have weak magnetic fields
and large filling factors, would seemlessly fit onto the regression curve for
the second population. This might be an indication that the two populations have
a certain spatial overlap. Thus, the flux density ratios could be a better
discriminator of the two populations rather than the heliocentric angle.

\section{Discussion and Conclusion\label{S-Discussion}}

In this study, we have exploited the SIR code's capability to simultaneously
handle many spectral lines using two independent model components inside the
resolution element so that it becomes possible to extract the height dependence
of many physical parameters. On the other hand, the SIR code works only in the
local thermodynamical equilibrium (LTE)
approximation, with plane-parallel atmospheres, and depth-independent filling
factors. This has to be considered in interpreting the results. Solar magnetic
fields are supposed to exist as kilo-Gauss magnetic elements (magnetic flux
tubes), which are distributed all over the solar surface. One place differs from
another only by the concentration of these elements (filling factor) inside the
resolution element (\opencite{Stenflo73}). Therefore, we expected at the outset
of our study that the SIR code would always provide kilo-Gauss magnetic fields
for the magnetic component with different but small filling factors depending on
the observed polarization signal. However, this assumption holds only for
locations, which are close to disk center. In contrast, closer to the limb, weak
magnetic fields with large filling factors and low temperatures are commonly
encountered.

The dependence of the two populations with weak and strong magnetic fields on
the heliographic angle points to the geometry of flux tubes as a possible
explanation. Features with high magnetic field strengths are related to small
filling factors. Thus, their horizontal extension must be small. Because of the
high field strengths, the gas pressure in these flux tubes is low, and they are
more transparent than their surroundings. They are heated through their walls
and appear hot. If these features are close to vertical with respect to the
solar surface, they are inclined to the LOS at a distance from disk center. An
observer sees only the higher part of these features, and close to the limb,
such features would be difficult to detect. In high photospheric layers, the
magnetic flux tubes must expand because of the lower outside pressure. In these
layers, the magnetic field has a horizontal component, which might contribute to
the circular polarization at locations away from the disk center, if the part
towards the observer has more weight than that on the opposite side. There are
two possible reasons for such a higher weight: (1) the optical path through the
line-forming layers might not be sufficiently long to also reach the opposite
side, or (2) the magnetic field lines are more (anti-)parallel to the LOS on
the side towards the observer than on the opposite side. Consequently, features
with low magnetic field strengths are more frequent closer to the limb, and they
are related to lower temperatures. This geometric interpretation is in agreement
with the findings of \inlinecite{Bommier} for internetwork magnetic fields.
However, we cannot exclude based on our observations that the flux tubes with
strong fields belong to the network.

Recently, \citeauthor{Stenflo2010} (\citeyear{Stenflo2010},
\citeyear{Stenflo2011}) also discovered two populations of weak and strong
magnetic fields in high-resolution \textit{Hinode} Fe\,\textsc{i} $\lambda
630.15$~nm and $\lambda 630.25$~nm magnetograms. The detection of these strong
and weak magnetic fields in two independent investigations with completely
different data sets is certainly of general interest and warrants further
scrutiny.

\begin{acks} We express our thanks to C. Denker for extended discussions
and many suggestions which significantly improved this article, and P. 
G\"om\"ory for carefully reading the manuscript.
This work was supported by the Deutsche Forschungsgemeinschaft (BA~1875/6-1).
MLD would like to express his appreciations to SOC/LOC of ESPM-13 for financial 
support which allowed him to attend the conference.
\end{acks}

\end{article}
\end{document}